\documentclass[aps,prb,twocolumn,amsmath,amssymb,showpacs]{revtex4-1}
\usepackage{graphicx}
\usepackage{dcolumn}
\usepackage{bm}

\begin{document}

\title{Spin-orbit interaction in a two-dimensional electron gas: 
a $SU(2)$ formulation}

\author{R.  Raimondi$^a$, P.  Schwab$^{b}$,  C.  Gorini$^c$, G.  Vignale$^d$}
\affiliation{$^a$CNISM and Dipartimento  di Fisica "E. Amaldi", Universit\`a  Roma Tre, 00146 Roma, Italy\\
$^b$Institut f\"ur Physik, Universit\"at Augsburg, 86135 Augsburg, Germany\\
$^c$Institut de Physique et Chimie des Mat\'eriaux de Strasbourg (UMR 7504),\\
CNRS and Universit\'e de Strasbourg, 23 rue du Loess, BP 43, 67034 Strasbourg Cedex 2, France\\
$^d$Department of Physics and Astronomy, University of Missouri, Columbia, Missouri 65211, USA
}

\date{\today}

\begin{abstract}
Spin-orbit interaction is usefully classified as extrinsic or intrinsic depending 
on its origin: the potential due to random 
impurities (extrinsic), or the crystalline potential associated with the band or device structure (intrinsic).
In this paper we will show how by using a $SU(2)$ formulation the two sources of spin-orbit 
interaction may be described in an elegant and unified way. 
As a result we obtain a simple description of the interplay of the two types of 
spin-orbit interaction,  and a physically transparent explanation of the vanishing of the d.c. spin Hall conductivity in a Rashba two-dimensional electron gas when spin relaxation is neglected, and its reinstatement when spin relaxation is allowed.  
 Furthermore,  we obtain an explicit formula for the transverse spin polarization created by an electric current, which generalizes the standard  formula obtained by Edelstein and Aronov and Lyanda-Geller by including extrinsic spin-orbit interaction and spin relaxation.
\end{abstract}
\maketitle                   

\section{Introduction: Three surprises}

Spin-orbit interaction in metals and semiconductors has attracted much interest in recent years
\cite{zutic2004, flatte2007}.
One reason for this is the possibility offered by such an interaction to achieve all-electrical
control of the electron spin, as e.g. in the case of the spin Hall effect.
The latter is a phenomenon in which a spin current appears in the direction transverse
to a charge current driven through a sample.
Our paper is motivated by several surprises and puzzles in the theory of the spin Hall effect 
in a two-dimensional electron gas (2DEG) {with linear Rashba (or Dresselhaus) spin-orbit coupling (this will be referred to in what follows as the {\em Rashba 2DEG})}.

The first surprise was Sinova and co-workers' observation that in the absence of disorder or interactions  the spin Hall conductivity assumes the universal value $\sigma^{sH} = e/8 \pi$\cite{sinova2004}. {This is in sharp contrast with the behavior of the regular electrical conductivity, which, of course,  is infinite under the stated assumptions, due to the absence of a scattering mechanism that may establish  a steady state regime.} The spin Hall conductivity -- within certain limitations -- 
 depends neither on the strength, nor on the microscopic form of the spin-orbit coupling.
 
The second surprise was that scattering from impurities, {while restoring a finite electrical conductivity}, leads to another
universal result, namely $\sigma^{sH}= 0$  for arbitrarily small impurity 
concentrations\cite{mishchenko2004,inoue2004,raimondi2005,khaetskii2006}.
This result can be traced back to a very specific form of the equation of motion 
for the spin polarization in the Rashba  2DEG, 
\begin{equation} \label{eq_0}
\partial_t s^y = - 2m \alpha j_{y}^z\,,
\end{equation}
where $s^y$ is the spin polarization in the $y$-direction, $m$ the effective electron mass in the sample, 
$\alpha $ the Rashba spin-orbit coupling constant 
and $ j_{y}^z$ the spin current polarized along $z$ and flowing along $y$\cite{rashba2004,dimitrova2005,chalaev2005}.
In a steady-state situation the time derivative of the spin polarization vanishes 
and so does the spin current and the d.c. spin Hall conductivity.

At this point a remark on the origin of spin-orbit coupling is in order. 
Spin-orbit interaction is usefully classified as extrinsic or intrinsic depending on its origin, 
either in the potential due to random impurities (extrinsic),  or in the crystalline potential associated with the band or device structure (intrinsic). 
In Refs. \onlinecite{engel2005} and \onlinecite{tse2006} the spin Hall effect was calculated 
for a 2DEG ignoring intrinsic spin-orbit coupling but assuming an  
extrinsic one of the form
\begin{equation}
H_{\rm so} = - \frac{ \lambda_0^2 }{4} {\boldsymbol \sigma}\times \nabla V(\mathbf{ x})\cdot \mathbf{ p} 
\end{equation}
where 
${\boldsymbol \sigma}=(\sigma_x,\sigma_y,\sigma_z )$ 
is the vector of Pauli matrices, $V(\mathbf{ x})$ is  the random impurity potential, and $\lambda_0$ the effective Compton wavelength characterizing the extrinsic spin-orbit interaction strength. 
Two distinct contributions to the spin Hall conductivity were identified, 
the skew scattering (ss) and side jump (sj) contributions:
$\sigma^{sH} = \sigma^{sH}_{\rm ss}+ \sigma^{sH}_{\rm sj}$,
with 
\begin{equation}
\label{ss0}
\sigma^{sH}_{\rm ss} =  \frac{\lambda_0^2}{16} e n p_F l m v_0  \, \text{ and }  \, \sigma^{sH}_{\rm sj}  = \frac{\lambda_0^2}{4} e n\,,
\end{equation}
where $n$ is the electron density, $p_F$ the Fermi momentum, $l$ the mean free path, and $v_0$ the amplitude of the 
impurity scattering potential.

{The next logical step for the theory was  to combine the two mechanisms, which of course are simultaneously present in experiments.  This led to a third surprise \cite{tse2006a,hankiewicz2008,cheng2008,raimondi2009}.  Namely, 
the spin Hall conductivity was found to be independent of the Rashba coupling constant $\alpha$, 
and yet its value  was different from the one at $\alpha = 0$.  In other words, the spin Hall conductivity appeared to be a {\em non-analytic} function of the Rashba coupling constant $\alpha$  -- its value exhibiting an unphysical jump from $\alpha=0$ to arbitrarily small but finite $\alpha$.} 
 
The issue of the non-analytic behavior of the spin Hall conductivity for $\alpha\rightarrow0$
was settled in Ref.~\onlinecite{raimondi2009}.  The ``unphysical jump" was shown to result from the  neglect of spin relaxation 
-- an effect that appears to {\em second order} in $\lambda_0^2$ and was therefore missed in the first-order 
treatments of Refs.~\onlinecite{tse2006a,hankiewicz2008,cheng2008}.  
When spin relaxation was included\cite{raimondi2009} the discontinuous jump of $\sigma^{sH}$ at  $\alpha=0$ was 
replaced by a smooth evolution on the scale of $\alpha= \frac{\hbar}{p_F\sqrt{\tau_s\tau}}$, where $1/\tau_s\, (1/\tau)$ is the spin (momentum) relaxation rate.

{However, an additional problem remains.  The theoretical results of Refs.~\onlinecite{tse2006a,hankiewicz2008,cheng2008,raimondi2009} 
agreed that the Rashba term suppresses the skew scattering contribution to the spin Hall conductivity for arbitrarily small $\alpha$. 
However, no consensus has been achieved so far on the 
side jump contribution.  While Refs.~\onlinecite{tse2006a} and \onlinecite{raimondi2009} claim   
$\sigma^{sH} = (1/2) \sigma^{sH}_{\rm sj}$, in Ref.~\onlinecite{hankiewicz2008} 
a vanishing spin Hall conductivity is predicted, $\sigma^{sH}=0$.}

Here we reconsider the problem in order to {settle} the issue of the side jump. 
We will make use of the analogy between spin-orbit coupling and a spin-dependent, 
i.e.~$SU(2)$, gauge field \cite{mathur1992, frohlich1993}.
We will start with a discussion of the equation of motion for the spin density which allows to determine 
the spin Hall conductivity in the presence of intrinsic {\it and} extrinsic mechanisms, with the result that $\sigma^{sH}=0$.
Subsequently we will also present the theory based on a generalized Boltzmann equation,
and will discuss in the appendices how this is microscopically derived from the Keldysh formalism.
{It will be shown how the simple dynamical argument for the vanishing of the spin Hall current in the steady state, Eq.~(\ref{eq_0}), is generalized to the more complex case in which both intrinsic and extrinsic sources of spin-orbit coupling are present.}

Finally, it is worth mentioning that quantum and semiclassical kinetic equations 
have been widely used to discuss spin-orbit coupling effects, 
see e.g.~\onlinecite{shytov2006, raimondi2006, sinitsyn2007, culcer2010, gorini2010}.  
Our treatment, which we here specialize to the Rashba model, is actually
valid for any system described by a Hamiltonian with linear-in-momentum
(intrinsic) spin-orbit coupling, possibly slowly varying in space and time;
as discussed in Ref.~\onlinecite{gorini2010}, a Zeeman field could also be included.
As such, it is a treatment which in principle allows one to study the interplay between
intrinsic and extrinsic spin-orbit coupling mechanisms in a wide range of systems.

\section{Equation of motion for the spin density}
In this section we will study the equation of motion for the spin density in a 2DEG including disorder, 
intrinsic and extrinsic spin orbit coupling with the Hamiltonian
\begin{equation}
\label{eq_1}
   H=\frac{  \mathbf{ p}^2}{2m}+\alpha ( p_y\sigma_x-p_x\sigma_y) +
V(\mathbf{ x})-\frac{\lambda_0^2}{4}{\boldsymbol \sigma}\times \nabla V(\mathbf{ x})\cdot \mathbf{ p}
.\end{equation}
Here $V({\bf x} )$ is the sum the potential created by the electric field that drives a current through the system
and a statistically fluctuating potential due to impurities, 
$V({\bf x} )= e {\bf E} \cdot {\bf x} + \delta V({\bf x} ) $ . 
The fluctuations of the impurity potential are given by
$\overline{ \delta V({\bf x} )\delta V({\bf x}')} = n_i v_0^2 \delta({\bf x} - {\bf x}')  $, 
$n_i$ being the impurity concentration and $v_0$ the scattering amplitude. 
In this paper we will assume weak spin-orbit coupling, in the sense that $\alpha p_F \ll \epsilon_F$ and
$ (\lambda_0 p_F)^2 \ll 1 $.
Since both spin-orbit coupling terms are linear in the momentum $ {\bf p } $ 
they can be parameterized by a $SU(2)$ vector potential,
\begin{equation}
   H = \frac{ {\bf p}^2 }{2m} + V({\bf x}) + \frac{1}{2m}  {\cal A}_i^a p_i \sigma_a
.\end{equation} 
Here the summation over indices $i$ and $a$ is implied. Like the scalar potential $V$ 
also the vector potential ${\cal A}_i^a$ is the sum of a smooth and a fluctuating contribution, 
${\cal A}_i^a = A_i^a + \delta A_i^a $, with
\begin{eqnarray}
\label{eq6}
A_i^a &=  &- 2 m \alpha \epsilon_{iaz} - \frac{1}{2}  e m  \lambda_0^2 \epsilon_{aji} E_j , \\
\label{eq7}
\delta A_i^a & = & - \frac{1}{2} m \lambda_0^2 \epsilon_{aji} \partial_j \delta V({\bf x} )
.\end{eqnarray}
The equation of motion for the spin density operator (operators will appear 
with a hat $\hat{ \hphantom{a} } $) assumes the compact form
\begin{equation}  \label{eqEOM}
\partial_t \hat s^a + \partial_i \hat j_i^a =  \epsilon_{abc}{\cal A}_i^b \hat j_i^c. 
\end{equation}
The spin density operator itself is defined as ($\hbar = 1$ )
\begin{equation}
\hat s^a({\bf x}) = \frac{1}{4} \left\{ \hat \sigma_a  ,  \delta({\bf x} - \hat {\bf r} ) \right\},
\end{equation}
whereas $\hat j_i^a$ are the components of the spin current operator. In analogy to the conventional current operator 
we write them as a sum of a paramagnetic and a diamagnetic term, $\hat j_i^a = [\hat j_P]_i^a + [\hat j_D]_i^a $, with
\begin{equation} \label{eq10}
[\hat j_P]_i^a =  \frac{1}{4m}  \left\{ \hat p_i \hat \sigma_a, \delta({\bf x} - \hat {\bf r} ) \right \},\quad 
[\hat j_D]_i^a  = \frac{1}{4m} {\cal A}_i^a \delta({\bf x} - \hat {\bf r})\,.
\end{equation}
The goal is to obtain the equation of motion for the (quantum and disorder) averaged spin density. 
While the l.h.s.~of Eq.~(\ref{eqEOM}) is easily averaged, the r.h.s.~involves the fluctuations 
of the vector potential and its correlations with the current density,
\begin{equation}
\partial_t s^a + \partial_i j_i^a = \epsilon_{abc} A_i^a j_i^a + \epsilon_{abc}  \overline{ \delta A_i^b \delta j_i^c}\,.
\end{equation}
We have not been able to determine the correlator on the r.h.s.~of the equation exactly.
An analysis carried out in the limits of weak spin-orbit coupling, weak disorder, and slow
spatial and temporal variations shows (see Appendix~\ref{AppA} for details) that 
\begin{equation}
\epsilon_{abc}  \overline{ \delta A_i^b \delta j_i^c}=0\,.
\end{equation}
More precisely, the correlations of the vector potential 
with the paramagnetic current operator cancel with the correlations of the diamagnetic current operator:
\begin{equation}
\epsilon_{abc} \overline{ \delta A_i^b [\delta j_P]_i^c } = - \epsilon_{abc} \overline{ \delta A_i^b  [\delta j_D ]_i^c }
 =  \frac{1}{2} \epsilon_{abc} A_i^b [ j_{\rm sj}]_i^c\,, 
\end{equation}
where  $[j_{\rm sj}]_i^c = \epsilon_{icj} \sigma^{sH}_{\rm sj} E_j$
is the spin Hall current from the side jump mechanism.
The averaged continuity equation is thus
\begin{equation} \label{eqContEq}
\partial_t s^a + \partial_i j_i^a = \epsilon_{abc} A_i^b j_i^c \,.
\end{equation} 
The argument for the vanishing spin Hall conductivity is thus the same 
as in the absence of extrinsic spin-orbit coupling: 
in a stationary and homogeneous situation the l.h.s.~of the equation is zero. 
From the $y$-component of (\ref{eqContEq}) we obtain $ 2m \alpha j_y^z =0$, and thus the spin Hall  conductivity is zero.


{We can gain a deeper  insight into the physical meaning of this  mathematical result by examining the structure of the Boltzmann 
equation in the presence of extrinsic and intrinsic $SU(2)$ vector potentials.  
We will show in the following sections that the net spin Hall current due to an electric field in the $x$-direction is the sum of

\begin{enumerate}
\item[(i)] A ``drift" contribution, associated with 
the skew-scattering and side-jump due to impurity scattering.

\item[(ii)] A Hall current due to the ``$SU(2)$ magnetic field" (see Section 3.2) associated with the $SU(2)$ vector potential of the Rashba spin-orbit coupling.

\item[(iii)]  A diffusion contribution, which is proportional to the  $y$-component of the uniform spin polarization.  
The spin polarization originates from the so-called {\it Edelstein effect}, i.e., 
the fact that electrons drifting in the $x$-direction in the presence of a Rashba field ``see" an effective magnetic field pointing 
in the $y$-direction.

\end{enumerate}

The presence of a spin diffusion current arising from a spatially homogeneous spin distribution  may appear puzzling at first,  but is a natural effect of the $SU(2)$ vector potential.  
The point is that the spin distribution, while spatially uniform, is  non-isotropic in spin space: 
therefore the spin-orbit coupling produces a spin diffusion current that tends to make the distribution isotropic.  Our results imply that the spin polarization $s^y$ created by the electronic current is significantly affected by spin-orbit coupling with the impurities, and differs from the standard Edelstein \cite{edelstein1990}  and Aronov and Lyanda-Geller \cite{aronov1989} prediction(see section 3.4).
}

%
%

\section{The Boltzmann equation}
In the following we will discuss the spin Hall effect in the context of the Boltzmann equation.
We  start with the standard Boltzmann equation in the absence of spin-orbit coupling, 
for which of course no spin Hall effect exists. 
In the subsequent steps intrinsic and extrinsic spin-orbit coupling will be included. 
In the extrinsic case we will
concentrate on the side jump, that is we will ignore skew scattering terms.
It will be demonstrated that in the presence of spin-orbit coupling a charge current 
generates a transverse spin current, i.e.~a spin Hall effect appears. 
In the Rashba model, however, a voltage induced spin polarization also appears,
whose contribution to the spin current exactly compensates the directly generated spin Hall current.
In other words, the very well known cancellation of the spin Hall effect in a (purely) Rashba 2DEG
is in fact more general, and holds even in the presence of extrinsic spin-orbit
coupling mechanisms.

We relegate technical details to the appendices, which can be skipped in a first reading.

\subsection{No spin-orbit coupling}
To fix the notation we begin by considering the Boltzmann equation when $\alpha =0$ and $\lambda_0=0$, 
i.e.~in the absence of spin-orbit interaction. We have
\begin{eqnarray}
&&\partial_t f_{\mathbf{ p}}+\frac{\mathbf{ p}}{m}\cdot \nabla_{\mathbf{ x}}f_{\mathbf{ p}}-e\mathbf{ E}\cdot \nabla_{\mathbf{ p}}f_{\mathbf{ p}} \nonumber\\
 &=&-2\pi n_i v_0^2 \sum_{\mathbf{ p}'}\delta (\epsilon_{\mathbf{ p}}-\epsilon_{\mathbf{ p}'})(f_{\mathbf{ p}}-f_{\mathbf{ p}'}),
 \label{eq_2}
\end{eqnarray}
with $\epsilon_{\mathbf{ p}}=\mathbf{ p}^2/2m$ and  $f_{\mathbf{ p}}\equiv f (\mathbf{ p}, \mathbf{ x},t)$ 
is the distribution function. In the following $f_{\mathbf{ p}}$ will be assumed to have a spin structure 
in order to describe charge and spin dynamics.
We follow the notation of Ref.~\onlinecite{gorini2010} and introduce the spin dependent density and current density as
\begin{equation}
\label{eq_3}
\rho =\sum_{\mathbf{ p}} f_{\mathbf{ p}}, \ \  {\mathcal J}=\sum_{\mathbf{ p}} \frac{\mathbf{ p}}{m} f_{\mathbf{ p}}.
\end{equation}
The particle density is given by $n={\rm Tr} (\rho)$,  
the spin density by $s^a={\rm Tr}(\sigma_a \rho)/2$, $a=x,y,z$, and
the particle and spin currents are defined analogously. 
Integrating the Boltzmann equation over the momentum $\mathbf{ p}$ gives the continuity equation
\begin{equation}
\label{eq_4}
\partial_t \rho +\nabla_{\mathbf{ x}}\cdot {\mathcal J}=0
,\end{equation}
i.e.~the continuity equations for the charge and spin components,
\begin{equation}
\partial_t n   + \nabla_{\mathbf{x} } \cdot {\mathbf j} =0, \quad
\partial_t s^a + \nabla_{\mathbf{x} } \cdot {\mathbf j}^a= 0
.\end{equation}

\subsection{Intrinsic spin-orbit coupling}
Next, we consider the Rashba spin-orbit interaction in the absence of extrinsic mechanisms, 
i.e.~we set $\lambda_0^2=0$ but allow $\alpha \ne 0 $. 
This case has been discussed several times in the literature. For a review one may see
Ref.~\onlinecite{vignale2010}. Here we follow the treatment of Ref.~\onlinecite{gorini2010}.
Using the $SU(2)$ formulation the Boltzmann equation becomes
\begin{eqnarray}
&&
\left(\partial_t +\frac{\mathbf{ p}}{m}\cdot \tilde{ \nabla}_{\mathbf{ x}}+\frac{1}{2}
 \left\{{\mathbf F}\cdot \nabla_{\mathbf{ p}} ,. \right\} \right) 
 f_{\mathbf{ p}}\nonumber \\
 & =& -2\pi n_i v_0^2 \sum_{\mathbf{ p}'}\delta (\epsilon_{\mathbf{ p}}-\epsilon_{\mathbf{ p}'})
  (f_{\mathbf{ p}}-f_{\mathbf{ p}'}).\label{eq_12}
\end{eqnarray}
Compared to the standard Boltzmann equation there are two modifications. 
First, in Eq.~(\ref{eq_12}), the spacial derivative becomes the covariant derivative expected 
for a $SU(2)$ vector potential ${\bf A}$ 
\begin{equation}
\label{eq_13}
\tilde{\nabla}_{\bf x} =\nabla_{\bf x} +{\rm i}  \left[ {\bf  A }, \dots \right],
\end{equation}
where $ [, ]$ denotes a commutator. Note that here the $SU(2)$ vector potential includes only the Rashba contribution, 
i.e.~only the first term on the r.h.s. of Eq.(\ref{eq6}).
The second modification concerns the force ${\bf F}$ which appears in the anticommutator $\{, \}$, 
since in addition to the electric field the force contains now a
spin-dependent $SU(2)$ Lorentz-like force,
\begin{equation}
\label{eq_14}
{\mathbf F}= -e{\bf  E}  - \frac{\mathbf p }{m} \times {\mathcal B}
, \quad  
{\mathcal B}_k=\frac{1}{2}\epsilon_{kln}{\rm i}  \left[ A_l,A_n\right].
\end{equation}
In the Rashba model the only non vanishing component of the ``$SU(2)$ magnetic field" is 
$ {\mathcal B}_z=-2 (m\alpha)^2 \sigma_z $. 
From the Boltzmann equation one finds immediately the continuity equation
\begin{equation}
\label{eq_16}
\partial_t\rho +\tilde{\nabla}_{\mathbf{ x}}\cdot {\mathcal J}=0,
\end{equation}
which has the same form as Eq.~(\ref{eq_4}) but for the spacial derivative, here replaced by the covariant one.
The covariant derivative does not change the equation for the density, but there appear extra terms in the
one for the spin density: the explicit result looks exactly like Eq.~(\ref{eqContEq}). 

\subsection{Extrinsic spin-orbit coupling}
In the next step we consider the so-called side jump effect due to extrinsic spin-orbit coupling,
while we neglect the intrinsic contribution, $\lambda_0^2 \ne 0$, $\alpha = 0 $. 
\begin{widetext}
The Boltzmann equation is modified and becomes
\begin{eqnarray}
 &  &\partial_t f_{\mathbf{ p}}+ \nabla_{\mathbf{ x}}\cdot \left[ \frac{\mathbf{ p}}{m} f_{\mathbf{ p}}+
2 \pi n_i v_0^2 \sum_{\mathbf{p}'} \delta(\epsilon_{\bf p} - \epsilon_{ {\bf p}'} ) 
 \frac{\lambda_0^2}{4}\frac{1}{2}\left\{ (\mathbf{ p }' - {\mathbf p}  )  
 \times {\boldsymbol \sigma} , f_{\mathbf{ p}'}\right\} \right]
-e\mathbf{ E}\cdot \nabla_{\mathbf{ p}}f_{\mathbf{ p}}  \nonumber\\
 & = &  -2\pi n_i v_0^2 \sum_{\mathbf{ p}'}\delta (\epsilon_{\mathbf{ p}}-\epsilon_{\mathbf{ p}'})
 \left[ 
 f_{\mathbf{ p}}-f_{\mathbf{ p}'}
 -\frac{\lambda_0^2}{4}\frac{1}{2} \left\{ (-e)\mathbf{ E}\cdot (\mathbf{ p}'-{\bf p}) \times {\boldsymbol \sigma},
 \frac{\partial f_{\mathbf{ p}'}}{\partial \epsilon_{\mathbf{ p'}}}\right\}
 \right].
\label{eq_5}
\end{eqnarray}
\end{widetext}
The extra terms due to spin-orbit coupling appearing in Eq.~(\ref{eq_5}) have been derived 
within the framework of a microscopic, Keldysh-based approach
along the lines of Refs.~\onlinecite{raimondi2009} and \onlinecite{raimondi2009a}.
More explicitly, the extra terms are obtained from the spin-orbit correction to the self-energy (Born approximation), 
compare Refs.~\onlinecite{raimondi2009,raimondi2009a} and Appendix~\ref{AppB}. 
The term on the l.h.s. of the equation appearing under the spacial derivative 
corresponds to a correction to the current operator, 
so that the  continuity equation (\ref{eq_4}) remains valid 
provided the expression for the current density is modified as follows 
\begin{equation}
\label{currentplus}
{\mathcal J} = \sum_{\bf p } \left[   \frac{\bf p }{m} f_{\bf p } + 
  \frac{\lambda_0^2}{4} \frac{1}{2\tau} \left\{ {\bf p } \times  { \boldsymbol \sigma } ,f_{\bf p } \right\} \right] 
.\end{equation}
We notice that both new terms contain the {\it side jump} shift 
$\Delta {\bf x}\equiv-(\lambda_0^2/4)\Delta {\bf p} \times {\boldsymbol\sigma}$, 
where $\Delta {\bf p}\equiv {\bf p'}-{\bf p}$. 
Such an expression of the side jump can be derived by analyzing the effect 
of the spin-orbit interaction on the scattering trajectory of a wave packet \cite{nozieres1973}.
 
Spin-orbit coupling is assumed to be small, and in the following we 
will calculate the spin Hall current to lowest order in the parameter 
$\lambda_0^2$, as usual in linear response to an homogeneous
applied electric field. Spin-orbit coupling generates a spin Hall current in two ways: 
(1) due to the modified expression for the current operator and (2) due to the spin-orbit correction to the 
distribution function $f_{\bf p }$. 

The spin-orbit correction to the current operator 
-- the second term on the r.h.s. of Eq.~(\ref{currentplus}) -- yields the spin current
\begin{equation}
\label{eq_6}
[j_{i}^{a}]^{(1)}=-\epsilon_{iab}\frac{\lambda_0^2}{8}\frac{m }{\tau} j_b,
\end{equation}
where $j_b $ is the particle current in the absence of spin-orbit coupling. 
Expressing the latter in terms of the Drude conductivity, $j_b = -(e n \tau/m) E_b$, the result reads
\begin{equation}
\label{eq_8}
[{j }_{i}^{a}]^{(1)}=\frac{1}{2}\frac{en\lambda_0^2}{4} 
\epsilon_{iab}E_b \equiv \epsilon_{iab} \frac{1}{2}\sigma_{\rm sj}^{sH} E_b.
\end{equation}
Eq.~(\ref{eq_8}) represents the first half of the side jump contribution to the spin Hall effect.

The second half is found when one considers the first term on the r.h.s.
of Eq.~(\ref{currentplus}) and computes the spin-orbit correction to the distribution function $f_{\bf p}$.
To this end, we multiply Eq.~(\ref{eq_5}) by $p_i/m$, 
integrate over the momentum and consider the $a$-th spin component.
\begin{widetext}
The last term on the l.h.s. of Eq.~(\ref{eq_5}) would give a term quadratic in the electric field,
therefore only the r.h.s. contributes and one is left with
\begin{equation}
0  =  -2\pi n_i v_0^2 \sum_{\mathbf{ p},\mathbf{ p}'} \delta (\epsilon_{\mathbf{ p}}-\epsilon_{\mathbf{ p}'})
        \frac{p_i}{m}\frac{{\rm Tr}(\sigma_a f_{\mathbf{ p}})}{2}
  -  2\pi n_i v_0^2 \sum_{\mathbf{ p},\mathbf{ p}'} \delta (\epsilon_{\mathbf{ p}}-\epsilon_{\mathbf{ p}'})
 \frac{\lambda_0^2}{4}\epsilon_{bkm}(-eE_b)\frac{\partial f_{\mathbf{ p}', eq}}{\partial \epsilon_{\mathbf{ p}'}}
 \frac{p_m p_i}{m}\frac{{\rm Tr}(\sigma_k \sigma_a)}{2}. 
\end{equation}
\end{widetext}
There follows
\begin{equation}
\label{eq_9}
[j_i^a]^{(2)} =  {\rm Tr} \sum_{\mathbf p } \frac{\sigma_a}{2}  \frac{\bf p}{m} f_{\bf p}
=\frac{1}{2}\frac{en\lambda_0^2}{4} \epsilon_{iab}E_b \equiv \epsilon_{iab} \frac{1}{2}\sigma_{\rm sj}^{sH} E_b.
\end{equation}
By combining Eqs.~(\ref{eq_8}) and (\ref{eq_9}) the total side jump contribution to the spin Hall conductivity 
reads $\sigma_{\rm sj}^{sH}= \frac{1}{4} en\lambda_0^2$.

\subsection{Intrinsic and extrinsic spin orbit coupling}
We are now ready to consider the interplay of intrinsic and extrinsic spin-orbit coupling, 
i.e.~both $\alpha$ and $\lambda_0^2$ are nonzero.
Since we assume weak spin-orbit coupling one could be tempted 
to write a Boltzmann equation in which the spin-orbit terms of (\ref{eq_5}) and (\ref{eq_12})
are simply added. 
Such an equation was obtained in Refs.~\onlinecite{raimondi2009,raimondi2009a}, 
however it does not correctly capture the side jump contribution the the spin Hall conductivity 
and leads to the wrong result $\sigma^{sH} = (1/2) \sigma^{sH}_{\rm sj}$.
Technically, what is missing in the derivation of the kinetic equation in \onlinecite{raimondi2009,raimondi2009a} 
(and also in the diagrammatic calculation of the spin Hall conductivity of \onlinecite{tse2006a}) 
is the modification to the self-energy
when intrinsic and extrinsic spin-orbit coupling are both present.
An advantage of the present $SU(2)$ approach is that such a correction appears quite naturally,
as discussed in Appendix \ref{AppB}. 
The bottom line of including such a modification is that the correct Boltzmann equation is obtained by replacing 
in Eq.~(\ref{eq_5}) the spacial derivative with the covariant one, 
the force with the sum of the standard electric force and the $SU(2)$ force, 
and by allowing for a correction to the collision integral 
due to the interplay of the two types of spin-orbit interaction.
In this way one obtains again the continuity equation (\ref{eq_16}) -- 
this time with $\mathcal{J}$ given by Eq.~(\ref{currentplus}) --
which implies the vanishing of the spin Hall conductivity. 

The diffusive regime allows to analyze how this happens in an analytically explicit way.
Let us consider first, for simplicity's sake, the pure Rashba case and set $\lambda_0 =0$.
We refer to 
\onlinecite{gorini2010} for details of the calculations.
The current density can be written as the sum of a diffusion, drift, and Hall current 
${\mathcal J } = {\mathcal J}_{\rm diff} + {\mathcal J}_{\rm drift} + {\mathcal J}_{\rm hall}  $
with
\begin{equation}
{\mathcal J}_{\rm diff }  = - D \tilde \nabla_{\bf x} \rho , \quad
{\mathcal J}_{\rm drift}  = \sigma {\bf E} , \quad
{\mathcal J}_{\rm hall }  = \frac{\tau}{2m} {\mathcal B } \times {\mathcal J}.
\end{equation}
Here $D = \frac{1}{2} v_F^2 \tau$ is the diffusion constant, and $\sigma$ the electrical conductivity.  Notice that the ``Hall conductivity" is taken from the classical formula for the weak magnetic field regime
$\sigma^{sH}_{int}=\sigma_D \omega_c\tau$,   where $\sigma_D$ is the Drude conductivity and $\omega_c$ is the cyclotron frequency associated with the ``$SU(2)$ magnetic field".
In the homogeneous case the explicit result for the spin current $j_y^z$ is 
\begin{equation}
\label{eq_18}
j_y^z=2m\alpha D s^y +\sigma_{\rm int}^{sH} E_x.
\end{equation}
The first term has its origin in the diffusion current, while the second term appears due to the Hall current.
One has
\begin{equation}
\label{eq_19}
\sigma_{\rm int}^{sH}=\frac{e}{8\pi}\frac{2\tau}{\tau_{DP}},
\end{equation}
where $\tau_{DP}$ is the Dyakonov-Perel spin relaxation time,  $1/\tau_{DP}=(2m\alpha)^2 D $.

Since we now know that, due to the continuity equation, the spin current $j_y^z$ must be zero, we can conclude that  there must exist a spin polarization in $y$-direction,
such that the diffusion current associated with it cancels the Hall current $\sigma_{\rm int}^{sH} E_x$.
Eq.~(\ref{eq_18}) then implies
\begin{eqnarray}
s^y & = & -\frac{1}{2m\alpha D}\sigma_{int}^{sH}E_x \nonumber\\
 & = & -e\alpha N_0 \tau E_x, \label{eq_21}
\end{eqnarray}
which is the well known result for the voltage induced spin polarization in the Rashba model\cite{edelstein1990,aronov1989}.
This may well be considered a new derivation of the Edelstein effect.  However, as we now show, the result for $s^y$ changes significantly when extrinsic contributions to the spin Hall current are taken into account.

When extrinsic spin-orbit interaction is present too, the spin Hall current in the diffusive limit becomes
\begin{equation}
j_y^z =  2m\alpha D s^y +(\sigma^{sH}_{\rm int}+\sigma_{\rm sj}^{sH})E_x\label{eq_24},
\end{equation}
so both the intrinsic and the extrinsic mechanisms drive a spin Hall current, 
but this is in the end compensated by the spin polarization $s^y$.
Notice that the vanishing spin Hall conductivity implies that the limit $\alpha \rightarrow 0$ 
does not reproduce the result obtained by setting $\alpha =0$ from the outset. 
This paradox has a simple solution as pointed out in \onlinecite{raimondi2009}: 
extrinsic spin-orbit coupling introduces the Elliott-Yafet spin relaxation mechanism,
which is order $\lambda_0^4$ and thus does not appear in our equations so  far. 
Elliott-Yafet spin relaxation is included by replacing Eq.~(\ref{eq_0}) by 
\begin{equation}
\label{eq_26}
\partial_t s^y + \frac{1}{\tau_s} s^y = - 2m\alpha j_y^z,
\end{equation}
with $1/\tau_s$ the Elliott-Yafet spin relaxation rate given by (see, for instance, Ref.~\onlinecite{raimondi2009a})
\begin{equation}
\label{eq_27}
\frac{1}{\tau_s}=\frac{1}{\tau}\left(\frac{\lambda_0 p_F}{2}\right)^4.
\end{equation}
From Eqs.~(\ref{eq_24}) and (\ref{eq_26}) one obtains
\begin{eqnarray}
j_y^z &=&\frac{1}{1+\tau_s /\tau_{DP}}(\sigma^{sH}_{\rm  int}+\sigma_{\rm sj}^{sH})E_x, \nonumber\\
s^y &=& - \frac{2 m \alpha}{1/\tau_s + 1/\tau_{DP}} ( \sigma^{sH}_{\rm  int}+\sigma_{\rm sj}^{sH})E_x
\label{eq_28}
.\end{eqnarray}
Eq.~(\ref{eq_28}) shows that the Elliott-Yafet spin relaxation indeed cures 
the non-analyticities in the limit $\alpha\rightarrow0$.

Finally, including the skew scattering contribution to the spin Hall current can be done by simply adding it
to Eq.~(\ref{eq_24}) to give 
\begin{equation}
j_y^z =  2m\alpha D s^y +(\sigma^{sH}_{\rm int}+\sigma_{\rm sj}^{sH}+\sigma_{\rm ss}^{sH})E_x\label{eq_24b}.
\end{equation}
For completeness in Appendix \ref{AppC} we discuss the physical origin 
of the skew scattering contribution and its microscopic evaluation. 
When using Eq.~(\ref{eq_24b}), rather than Eq.~(\ref{eq_24}), into Eq.~(\ref{eq_26}), one obtains
the same expressions as in Eq.~(\ref{eq_28}) with the replacement 
$\sigma_{\rm sj}^{sH}\rightarrow \sigma_{\rm sj}^{sH}+\sigma_{\rm ss}^{sH}$.
Observe that the expression ~(\ref{eq_28}) for $s^y$ is significantly different from the standard  expression~\cite{edelstein1990,aronov1989}.  The standard expression is recovered only in the limit in which  $1/\tau_s \to 0$ and the extrinsic contribution to $\sigma^{sH}$ is neglected.

\section{Summary}

The interplay between intrinsic and extrinsic spin-orbit coupling effects in 2DEGs
is an experimentally relevant issue whose theoretical description
has proven rather delicate and riddled with puzzles.
Moreover, the latter has suffered from the lack of a physically transparent formulation.

We have shown how formulating spin-orbit coupling in terms of non-Abelian gauge fields
can be exploited to unify the description of intrinsic and extrinsic effects.
Though for clarity's sake, and to establish a direct contact with previous works,
we have focused on the Rashba model and thus on $SU(2)$ fields,
our approach can be used for any Hamiltonian with a linear-in-momentum 
intrinsic spin-orbit term -- possibly slowly varying in space and time, too.
In such a unified picture the spin-charge coupled dynamics 
is naturally described by a set of simple continuity equations.
In particular, this allowed us to clarify the issue of the side jump contributions
to the spin Hall effect in the presence of Rashba spin-orbit interaction.
Furthermore, we have derived a new formula, Eq.~(\ref{eq_28}), for the transverse spin polarization associated with the electric current.  This is significantly different from the  Edelstein and Aronov and Lyanda-Geller formula and is amenable to experimental verification.

We acknowledge useful discussions with Dimitrie Culcer.
This work has been supported by the Deutsche Forschungsgemeinschaft through SPP1285
and by the French Agence Nationale de la Recherche through grant No.~ANR-08-BLAN-0030-02.
GV acknowledges support from NSF Grant DMR-0705460.
CG acknowledges the hospitality and support of the University of Missouri
where part of this work was done.

%

\appendix
\section{Equation of motion for the spin density -- some technical details}\label{AppA}
In the main text we showed that the equation of motion for the spin density can be written as
\begin{equation}
\partial_t s^a + \partial_i j_i^a = \epsilon_{abc} A_i^b j_i^c +
         \epsilon_{abc} \overline{\delta A_i^b \delta j_i^c }  
.\end{equation}
We mentioned that the correlations between the vector potential and the current vanish 
without giving any details.  Some will be given in this appendix.
\begin{widetext}
We begin by noting that with the introduction of the covariant momentum $\Pi_i$ defined by
\begin{equation}
\label{a_1}
\Pi_i =-{\rm i}\partial_i +\frac{1}{2} A_i^b\sigma_b
\end{equation}
the paramagnetic and diamagnetic currents can be combined together to give for the total current
\begin{equation}
\label{a_2}
\delta j_i^c ({\bf x}) =\frac{1}{8m}{\rm Tr} \sum_{{\bf p},{\bf p'}} 
\left\{ (\Pi_i +{\Pi'}_i),\sigma_c\right\} \langle c^{\dagger}_{\bf p}c_{\bf p'}\rangle  e^{{\rm i}({\bf p'}-{\bf p})\cdot {\bf x}},
\end{equation}
where the density matrix $\langle c^{\dagger}_{\bf p}c_{\bf p'}\rangle$ is a matrix in spin space.

The fluctuating vector potential can be written as
\begin{equation}
\label{a_3}
\delta A_i^b({\bf x})=
{\rm i}\frac{m\lambda_0^2}{2}\epsilon_{ijb}\sum_{\bf q} q_j \delta V({\bf x}) e^{{\rm i}{\bf q}\cdot {\bf x}}.
\end{equation}

We then need to evaluate the average over impurities
\begin{equation}
\label{a_4}
\overline{\delta A_i^b({\bf x})\delta j_i^c ({\bf x})}=
{\rm i}\frac{\lambda_0^2}{8}{\rm Tr} \sum_{{\bf p},{\bf p'}} 
\left\{ ({\boldsymbol\Pi}\times {\boldsymbol \Pi}')_b ,\sigma_c\right\}
\overline{\delta V ({\bf p}-{\bf p'})\langle c^{\dagger}_{\bf p}c_{\bf p'}\rangle}.
\end{equation}
In obtaining the above equation, we have used the fact that the standard white-noise disorder correlations,
$\overline{\delta V({\bf x})\delta V({\bf x'})}=n_i v_o^2 \delta ({\bf x}-{\bf x'})$, 
require $\overline{\delta V({\bf q})\delta V({\bf q'})}=n_iv_0^2 \delta ({\bf q}+{\bf q'})$, 
so that $q_j$ in Eq.~(\ref{a_3}) must be equal to $p_j-{p'}_j$, 
which in turn is equal to $\Pi_j -{\Pi '}_j$. We may then use the identity
$$
\epsilon_{ijb}q_j\left\{ \Pi_i+{\Pi '}_i,\sigma_c\right\}=
2\left\{ ({\boldsymbol\Pi}\times {\boldsymbol \Pi}')_b ,\sigma_c\right\}
$$
and obtain Eq.~(\ref{a_4}).

To perform the average in Eq.~(\ref{a_4}) we write the density matrix in terms of the Keldysh Green function to read
\begin{equation}
\label{a_5}
\overline{\delta A_i^b({\bf x})\delta j_i^c ({\bf x})}
=
-{\rm i} n_iv_0^2\frac{\lambda_0^2}{8}{\rm Tr} \sum_{{\bf p},{\bf p'}} \int \frac{{\rm d}\epsilon}{4\pi {\rm i}}
 \left\{ ({\boldsymbol\Pi}\times {\boldsymbol \Pi}')_b ,\sigma_c\right\}
\left( G_{\bf p'}^R (\epsilon )G_{\bf p}^K (\epsilon )+G_{\bf p'}^K (\epsilon )G_{\bf p}^A (\epsilon )\right).
\end{equation}
\end{widetext}

Then, by observing that to first order in ${\bf  p}_F  \cdot {\bf  A } / (m \epsilon_F) $
\begin{equation}
\label{a_6}
\sum_{\bf p}{\boldsymbol\Pi}  \ {\cal I}m \ G^{R}_{\bf p}(\epsilon )=0,
\end{equation}
the cancellation mentioned in the main text follows.

\section{Side jump corrections in the Boltzmann equation}
\label{AppB}
We determine the Boltzmann equation for a system with intrinsic and extrinsic spin-orbit coupling following the procedure of \onlinecite{gorini2010}.
The starting point is the Dyson equation for the (nonequilibrium) Green function
from which the equation of motion 
\begin{equation}  \label{eq47}
\left[ \partial_t + \frac{\bf p}{ m } \cdot \tilde \nabla_{\bf x} 
    - e \frac{\bf p }{m  } \cdot {\bf E} \, \partial_\epsilon 
    + \frac{1}{2} \{ {\bf F}, \nabla_{\bf p } \cdot  \} 
\right] \tilde G =
-i [ \tilde \Sigma, \tilde G ] 
\end{equation}
for the ``locally covariant'' Green function $\tilde G$ is obtained. 
$t$ and  ${\bf x}$ are the center-of-mass coordinates of the 
two-point Green function $\tilde G$ and $\epsilon$, ${\bf p}$ are the Fourier transformed relative coordinates.
Furthermore $\tilde G$ and $\tilde \Sigma$ have a structure in Keldysh space, 
i.e.~they have retarded, advanced and a Keldysh components.
The locally covariant Green function, $\tilde G$, is related to the conventional Green function $G$ by the transformation
\begin{equation}
\label{b_2}
\tilde G({\bf p}) =  G({\bf p}) - \frac{1}{2} \{ {\bf A}\cdot\nabla_{\bf p }, G({\bf p } ) \}
,\end{equation}
where ${\bf A}$ is the $SU(2)$ vector potential.
The distribution function appearing in the Boltzmann equation is given 
by the Keldysh component integrated over the energy,
\begin{equation}
f_{\bf p } = \frac{1}{2} \left[ 1 + \int \frac{d \epsilon}{ 2 \pi i}  \tilde G^K({\bf p}, \epsilon)  \right]
,\end{equation}
so that by integrating (\ref{eq47}) over the energy one obtains the Boltzmann equation for the distribution function $f_{\bf p}$.

Generally, the r.h.s. of (\ref{eq47}) is responsible for the scattering kernel in the Boltzmann equation.
For example, within the Born approximation and for spin-independent scattering the self-energy is
\begin{equation}
\tilde \Sigma = n_i v_0^2 \sum_{\bf p } \tilde G({\bf p })
\end{equation}
and leads to the r.h.s. of Eq.~(\ref{eq_12}).

The side jump corrections are found from the Born self-energy to first order in the (extrinsic) spin-orbit  
coupling, $\lambda_0^2$. 
The expression for the self-energy is given in Ref.~\onlinecite{raimondi2009a} for  $\alpha = 0$. 
In the presence of Rashba spin-orbit coupling the covariant expression, 
obtained by using the transformation (\ref{b_2}), reads 
$\tilde\Sigma^1 = \tilde\Sigma^1_a + \tilde\Sigma^1_b + \tilde\Sigma^1_c$ with 
\begin{eqnarray}
\tilde\Sigma^1_{a}({\bf p}, {\bf x} )  & = &  {\rm i} \frac{\lambda_0^2}{ 4}  n_i v_0^2 
  \epsilon_{abc} \sum_{{\bf p}' }  \left[{ p}_a  \sigma_b {p'_c} ,
    \tilde G({\bf p}', {\bf x}) \right],\label{sjpotential} \\
\tilde\Sigma^1_{b}({\bf p}, {\bf x} )  & = & \frac{\lambda_0^2}{8} n_i v_0^2
  \epsilon_{abc } \sum_{{\bf p}'}\lbrace  \sigma_b p'_c,(\tilde\nabla_{\bf x})_a \tilde G({\bf {p}'},{\bf x})\rbrace 
 \label{sjcurrent},\\
\tilde\Sigma^1_{c}({\bf p}, {\bf x}) & = & -\frac{\lambda_0^2}{8} n_i v_0^2
 \epsilon_{abc} \sum_{{\bf p}'}(\tilde\nabla_{\bf x})_a\lbrace {p}_c \sigma_b ,
  \tilde G({\bf p}',{\bf x})\rbrace
 \label{sjelectric}
,\end{eqnarray}
where $(\tilde\nabla_{\bf x})_a$ is the (spin) $a$-component of
the covariant derivative defined in Eq.(\ref{eq_13}) in the main text. 
Notice that the covariant derivatives appearing in the self-energies
$\tilde\Sigma^1_b$ and $\tilde\Sigma^1_b$ act once inside and once outside the anticommutator.
This is because the vector potential $A_a$ does not commute with the Pauli matrices $\sigma_b$. 
When $\alpha =0$, the covariant derivative reduces to the simple spacial derivative 
and Eqs.~(\ref{sjpotential}-\ref{sjelectric}) reduce to the expressions reported in Ref.~\onlinecite{raimondi2009a}.
In this case both $\tilde\Sigma^1_b$ and $\tilde\Sigma^1_c$ can be written as divergence terms 
and taken to the l.h.s. of the kinetic equation (\ref{eq47}), giving rise to the 
spin-orbit corrections appearing on the l.h.s. of Eq.~(\ref{eq_5}). 

In the presence of an electric field the space derivative has to be replaced by
\begin{equation}
\label{b_10}
\nabla_{\bf x}  \to \nabla_{\bf x} - e {\bf E} \partial_\epsilon
,\end{equation}
which gives rise to the electric field-dependent term on the r.h.s. of Eq.(\ref{eq_5}).

Finally, we note that although $\tilde\Sigma^1_{a}$ is the leading term, it is 
not relevant for the side jump contribution the to spin Hall effect and has therefore been
ignored in Eq.~(\ref{eq_5}). 
Indeed, this term is responsible for the so-called swapping of the spin currents\cite{lifshits2009}.

When Rashba spin-orbit coupling is present, inserting the self-energies (\ref{sjpotential}-\ref{sjelectric})
into the kinetic equation (\ref{eq47}) leads to a Boltzmann equation where the side jump corrections to the current
appear under the covariant derivative, confirming the covariant form of the continuity equation discussed in Section 2.
In addition, on the r.h.s. of the Boltzmann equation the additional term
\begin{equation}
\label{b_11}
\frac{1}{\tau}\frac{\lambda_0^2}{8}\sum_{\bf p'}\delta (\epsilon_{\bf p}-\epsilon_{\bf p'})\epsilon_{abc}\left\{ (\tilde\nabla_{\bf x})_a \sigma_b , p_cf_{\bf p}-{p'}_cf_{\bf p'}\right\}
\end{equation}
appears.
This can be interpreted as a correction to the collision integral 
arising from the interplay of extrinsic and intrinsic spin-orbit coupling. 
Such term, which is also manifestly covariant, is important for the consistency of the kinetic equation. 
However it can be easily seen that, due to the symmetry between ${\bf p}$ and ${\bf p'}$,
it vanishes after summation over ${\bf p}$ and hence does not modify the continuity equation.

\section{Skew scattering}
\label{AppC}
For completeness we now discuss skew scattering.
Let us assume a system with time reversal symmetry, but spin dependent scattering due to spin-orbit interaction.
The scattering amplitude reads
\begin{equation}
\label{ss1}
S=A+{\hat {\bf p}}\times {\hat {\bf p}'}\cdot {\boldsymbol \sigma} B,
\end{equation}
where ${\hat {\bf p}}$ and ${\hat {\bf p}'}$ are unit vectors in the direction of the momentum before and after the scattering event.
By considering the scattering probability proportional to $|S|^2$, one obtains three contributions given by $|A|^2$, $|B|^2$ and
$2 \ {\cal R}e \  (A B^*) {\hat {\bf p}}\times {\hat {\bf p}'}\cdot {\boldsymbol \sigma}$. 
Whereas the first two contributions are spin independent and give the total scattering time, 
the third one represents the so-called skew scattering term according to 
which electrons with opposite spin are scattered in opposite directions. 
To lowest order in perturbation theory or Born approximation one has 
$A=v_0$ and $B=-{\rm i}(\lambda_0^2p_F^2 /4) v_0$, having adopted a point-like impurity potential 
$V({\bf x})=v_0\delta ({\bf x})$.
Clearly, since $A$ and $B$ are out of phase, there is no skew scattering effect to this order. 
For it to appear to first order in the spin-orbit coupling constant $\lambda_0^2$, 
$A$ has to be evaluated beyond the Born approximation. 
The scattering problem can be cast in terms of the Lippman-Schwinger equation
\begin{equation}
\label{ss2}
\psi ({\bf x}) =e^{{\rm i}{\bf k}\cdot {\bf x}}+\int {\rm d}{\bf x}' \ G ({\bf x}-{\bf x}')V({\bf x}')\psi ({\bf x}'),
\end{equation}
where $G({\bf x})$ is the retarded Green function at fixed energy.
From (\ref{ss2}) we get
\begin{eqnarray}
\psi^{(1)}=v_0G({\bf x}), && A^{(1)}=v_0;  \nonumber\\
\psi^{(2)}=v_0^2 G({\bf 0}) G({\bf x}), && A^{(2)}=v_0^2G({\bf 0}).
\label{ss3}
\end{eqnarray}
Notice that only the imaginary part of $A^{(2)}$ is needed. By recalling that ${\cal I}m \  G({\bf 0})=-\pi N_0$, 
the spin-orbit independent scattering amplitude reads
\begin{equation}
\label{ss4}
A=v_0 \left( 1-{\rm i}\pi N_0 v_0\right).  
\end{equation}
The skew scattering contribution will then follow by inserting the modified scattering amplitude (\ref{ss4})
into the collision integral of the Boltzmann equation.
The same result can, of course, be obtained in quantum field theory using the Green function technique. 
The latter becomes necessary when one wants to consider skew scattering 
in the presence of Rashba spin-orbit interaction. 
To this end one has to consider the electron self-energy at least to third order in the scattering potential $v_0$.  
The expression for this was obtained in Ref.~\onlinecite{raimondi2009a} for the case $\alpha =0$ 
and gives rise to three terms depending on the position of the insertion of 
$- (\lambda_0^2/4) {\boldsymbol \sigma}\times \nabla V({\bf x})\cdot {\bf p}$ into a third order diagram. 
When Rashba spin-orbit coupling is present 
one has to consider the covariant self-energy, as done for the side jump contribution in appendix \ref{AppB}. 
It may be shown that to leading order, this is done simply by replacing the Green function $G$ 
with its covariant expression $\tilde G$. Hence the self-energy responsible for the skew scattering reads
$\tilde \Sigma^{\rm ss}=\tilde \Sigma_a^{\rm ss}+\tilde \Sigma_b^{\rm ss}+\tilde \Sigma_c^{\rm ss}$, where
\begin{eqnarray}
\tilde \Sigma_a^{\rm ss} & = & -{\rm i} n_i v_0^3\frac{\lambda_0^2}{4} \sum_{{\bf p}_a,{\bf p}_b} \  \tilde G({\bf p}_a) \  \tilde G({\bf p}_b)  \ {\bf p}_b \times {\bf p}\cdot {\boldsymbol \sigma} \label{ss5}\\
\tilde\Sigma_b^{\rm ss} & = &  -{\rm i} n_i v_0^3\frac{\lambda_0^2}{4}\sum_{{\bf p}_a,{\bf p}_b}\  \tilde G({\bf p}_a) \ {\bf p}_a \times {\bf p}_b \cdot {\boldsymbol \sigma} \  \tilde G({\bf p}_b)  \ \label{ss6}\\ 
\tilde \Sigma_c^{\rm ss} &=&   -{\rm i} n_i v_0^3\frac{\lambda_0^2}{4} \sum_{{\bf p}_a,{\bf p}_b} \ {\bf p} \times {\bf p}_a \cdot {\boldsymbol \sigma} 
\  \tilde G({\bf p}_a) \  \tilde G({\bf p}_b) \label{ss7}.
\end{eqnarray}
Here the standard impurity average has been performed, so that translational invariance is recovered
and the self-energy depends on the external momentum ${\bf p}$ only. 
Since we are considering the effect to first order in $\lambda_0^2$,
the covariant Green functions entering Eqs.(\ref{ss5}-\ref{ss7}) are spin independent and isotropic in momentum space. 
As a result the retarded and advanced components of the above self-energies vanish, 
while the Keldysh component survives only for $\tilde \Sigma_a^{\rm ss}$ and $\tilde \Sigma_c^{\rm ss}$. 
Their joint contribution, after recalling that $\sum_{{\bf p}}\tilde G^R({\bf p})=-{\rm i}\pi N_0$, 
leads to an extra term on the r.h.s. of the Boltzmann equation
\begin{equation}
\label{ss8}
-2\pi n_i v_0^2  (v_0 \pi N_0) \frac{\lambda_0^2}{4}\sum_{{\bf p'}}\delta (\epsilon_{\bf p}-\epsilon_{\bf p'}) \left\{   {\bf p'}\times {\bf p}\cdot {\boldsymbol \sigma}, \ f_{\bf p'}\right\}.
\end{equation}
Its contribution to the spin current is obtained as usual 
by multiplying Eq.~(\ref{ss8}) by $p_i/m$, integrating over momentum and projecting on the $a$-th spin component.
Again, (\ref{ss8}) is already order $\lambda_0^2$, so the distribution function $f_{\bf p'}$ is the one
in the absence of spin-orbit coupling.
\begin{widetext}
One has
\begin{equation}
-2\pi n_i v_0^2  (v_0 \pi N_0) \frac{\lambda_0^2}{4}\sum_{{\bf p'},{\bf p}}\delta (\epsilon_{\bf p}-\epsilon_{\bf p'}) 
\frac{p_i}{m}\epsilon_{jlk}  {p'}_j  { p}_l \frac{{\rm Tr}( { \sigma}_k\sigma_a)}{2} \ 2 f^0_{\bf p'}
=\epsilon_{iaj}\frac{1}{\tau}(v_0 \pi N_0)\frac{(\lambda_0p_F)^2}{8}\frac{en\tau}{m}E_j,
\label{ss8b}
\end{equation}
and the expression appearing in Eq.~(\ref{ss0}) follows at once.
\end{widetext}



 \bibliography{paper}

\end{document}